\documentstyle[preprint,aps]{revtex}
\newcommand{\be}{\begin{equation}}
\newcommand{\ee}{\end{equation}}
\newcommand{\bea}{\begin{eqnarray}}
\newcommand{\eea}{\end{eqnarray}}
\begin{document}
\draft

\title{Secondary Instabilities and Spatiotemporal Chaos in Parametric
Surface Waves}

\author{Wenbin Zhang and Jorge Vi\~nals}

\address{ Supercomputer Computations Research Institute, Florida State
University \\
Tallahassee, Florida 32306-4052}

\date{\today}
\maketitle

\begin{abstract}
A two dimensional model is introduced to study pattern
formation, secondary instabilities and the transition to spatiotemporal chaos
(weak turbulence) in parametric surface waves. The stability of a
periodic standing wave state above onset is studied against Eckhaus,
zig-zag and transverse amplitude modulations (TAM) as a function of the control
parameter $\varepsilon$ and the detuning. A mechanism
leading to a finite threshold for the TAM instability is identified.
Numerical solutions of the model
are in agreement with the stability diagram, and also reveal the
existence of a transition to spatiotemporal chaotic states at a finite
$\varepsilon$. Power spectra of temporal fluctuations in the chaotic state
are broadband, decaying as a power law of the frequency $\omega^{-z}$ with
$z \approx 4.0$.

\end{abstract}

\vspace{0.5in}

\narrowtext

A layer of incompressible fluid that is driven by a sinusoidal force
normal to the free surface at rest exhibits parametrically excited
surface waves, also known as Faraday waves \cite{re:mi90}. We introduce a two
dimensional model that possesses many of the
features observed in experimental studies of Faraday waves: a primary
instability to a standing wave pattern, a secondary instability (transverse
amplitude modulation or TAM) at larger amplitudes of the dimensionless driving
force $\varepsilon$, and a chaotic state at yet larger values of
$\varepsilon$.
Our study focuses on the large aspect ratio limit in which
the wavelength of the wave is much smaller than the lateral dimension of
the fluid layer.

Transitions to spatiotemporal chaos have been observed in a variety of
physical systems, with recent efforts concentrating on the study of
fluid systems \cite{re:cr93}. The fact that physical properties of fluids
are normally well
characterized, and that in some cases the transition to chaos occurs at
reasonably small
values of the control parameter, makes fluid systems especially
attractive for an accurate
quantitative study of the chaotic state. Among these recent studies we quote
rotating convection in a circular geometry \cite{re:ni93}, spiral defect
chaos in Rayleigh-B\'enard convection \cite{re:mo93}, and Faraday waves
\cite{re:tu89}. Theoretical work, on the other hand,
has concentrated on the study of
coupled lattice maps \cite{re:ka89,re:gr91,re:bo92}, in which simple maps
that are known to be chaotic are arranged on a lattice and their
dynamics coupled. Complementary studies focus on phenomenological models
such as the complex Ginzburg-Landau equation \cite{re:cr93}. This equation
models the low frequency behavior of a class of systems
close to onset of instability, and
exhibits various types of
chaotic behavior. A third approach, which we follow here, concerns the
so called order parameter equations. The best known example of the
latter is the Swift-Hohenberg equation used to model Rayleigh-B\'enard
convection in a Boussinesq fluid \cite{re:sw77}. Its associated
amplitude equation, the real Ginzburg-Landau equation, coincides with the
amplitude equation directly derived from the equations governing fluid motion.
The Swift-Hohenberg equation,
supplemented with non-gradient terms, has been recently used to study
spiral defect chaos in Rayleigh-B\'enard convection \cite{re:xi93}.

Recent experiments on Faraday waves in large aspect ratio systems have
revealed a number of interesting phenomena
\cite{re:tu89,re:ez86,re:ci91,re:ch92,re:bo93,re:gl93}. Among them,
periodic standing wave patterns near onset are found to be unstable
against a TAM at some finite supercriticality \cite{re:tu89,re:ez86,re:ch92}.
Associated with this
instability, temporal fluctuations of the pattern have been observed,
with a characteristic time scale that decreases continuously with
increasing $\varepsilon$. Beyond a certain value of $\varepsilon$,
the wave patterns appear temporally chaotic and
spatially disordered \cite{re:tu89,re:bo93}.
In the disordered regime, the amplitudes of a range of
Fourier modes appear to exhibit Gaussian statistics \cite{re:tu89,re:ho89}
(deviations from Gaussianity have also been recently reported
\cite{re:bo93}). Interestingly,
disordered wave patterns in the chaotic regime
have been found to exhibit highly ordered time averages \cite{re:gl93}.
The origin of the TAM instability and its relationship with the
disordered state was first studied theoretically
by Ezerskii et al. \cite{re:ez86}, who
derived a one
dimensional model to describe the modulation. Later, Milner \cite{re:mi91}
derived a set of coupled amplitude equations for a two dimensional surface,
including nonlinear interaction and damping of the waves. He showed that
a pattern of square symmetry is realized near onset in fluids of low
viscosity, and that it can become unstable against a TAM.
However, its finite threshold and its
dependence on the detuning parameter are not yet well understood
\cite{re:cr93}. Studies of chaotic wave patterns were also conducted by
Rabinovich and coworkers \cite{re:ra90}, who numerically studied an
amplitude equation
for a pair of counter-propagating waves, and found features
similar to the chaotic states observed in the experiments.

The model that we have derived for a complex order parameter $\psi$
preserves the rotational symmetry of the fluid, reduces to Milner's
amplitude equations when $\psi$ is assumed to be modulated along
particular directions of propagation, and to Ezerskii et al. one
dimensional equation in the case they considered. We confine ourselves
in this paper to the
results obtained when only the simplest nonlinear term is
retained in the model. With this choice of
nonlinearity we find a standing roll wave pattern close to onset.
The exact form of the nonlinear term in momentum
space, and the
results corresponding to different choices of nonlinearity that yield
standing waves of square symmetry
at onset will be presented elsewhere \cite{re:zh94a}.
We then obtain the stability boundaries of the roll pattern for either phase
(Eckhaus and zig-zag) or amplitude (TAM) instabilities.
A small but finite nonlinear damping coefficient appears to be
essential for the existence of the finite driving
threshold for the TAM instability. Unfortunately, and similarly to previous
studies, we do not have a self-consistent procedure to introduce
nonlinear damping into the model, probably reflecting the fact that at this
level it has to be regarded as phenomenological.
Numerical integration is finally
used to study pattern selection above onset, and a transition to spatiotemporal
chaotic states mediated by a TAM at finite $\varepsilon$.

The model equation that we study in this paper is,
\be
\label{eq:model_eq}
\partial_{t}\psi = -\gamma \psi + if\psi^{*}
+ \frac{3i}{4}(1+\nabla^{2})\psi + (i-\gamma_n) |\psi|^2\psi,
\ee
where $\psi(x,y,t)$ is a two-dimensional complex field. Its real and
imaginary parts are linear combinations of the surface displacement away from
planarity and the surface velocity potential of the irrotational part of the
flow \cite{re:zh94a,re:za68}.
Equation (\ref{eq:model_eq}) has been made dimensionless by choosing
$1/q_0$ as a unit of length, and $2/\Omega$ as a unit of time ($\Omega$ is
the angular frequency of the sinusoidal driving force, and $q_0$ is critical
wavenumber at onset, determined by both $\Omega$ and the
surface tension of the fluid). The linear damping coefficient is
$\gamma = 4\nu q_0^2/\Omega$, where $\nu$ is the kinematic viscosity
of the fluid.  The nonlinear damping coefficient $\gamma_n$ is of the order of
$\gamma$, but its exact value is undetermined.
The quantity $\varepsilon = (f-\gamma)/\gamma$ is the distance away from
threshold of the primary instability, where $f$ is the dimensionless
amplitude of the driving force.
The phase $e^{-i\Omega t/2}$ in the motion of the surface has
been scaled out of the field $\psi(x,y,t)$. Equation (\ref{eq:model_eq}) can be
derived perturbatively from the fluid equations in the limit of weak viscous
dissipation and small wave-steepness \cite{re:zh94a}.
With the nonlinear term used in Eq.~(\ref{eq:model_eq}),
standing roll wave patterns are observed just above $\varepsilon = 0$.
We believe, however, that our findings about secondary
instabilities, and the general features of the transition to spatiotemporal
chaos are qualitatively similar to the more realistic case in which the
standing
wave pattern near onset exhibits square symmetry.

For $\varepsilon > 0$, the quiescent solution $\psi = 0$ loses stability. A
new stationary roll solution can be found approximately by considering a
one-mode Galerkin approximation $\psi_0(x,q) = \alpha_q e^{i\theta_q}\cos(qx)
+ {\cal O}(\alpha_q^3)$ with,
$$
\alpha_q^2 =
\frac{q^2-1-4\gamma\gamma_n/3 + \sqrt{(q^2-1-4\gamma\gamma_n/3)^2
      + (1+\gamma_n^2)\left[16(f^2-\gamma^2)/9-(q^2-1)^2\right]}}{
     1+\gamma_n^2},
$$
$\sin 2\theta_q = (\gamma+3\gamma_n\alpha_q^2/4)/f$, and
$\cos 2\theta_q = \frac{3}{4}(q^2-1-\alpha_q^2)/f$.
Another solution, which does not exist for $q=1$ and is unstable with respect
to
uniform amplitude perturbations, will not be considered here. The neutral
stability
curve is $\varepsilon_0(q) = \sqrt{1+\left[3(1-q^2)/(4\gamma)\right]^2}-1$.
The roll solution breaks the translational invariance of the base state. Hence
a phase
diffusion equation can be derived to study the stability of the roll solution.
We
find that the Eckhaus stability boundary is given by \cite{re:zh94b},
\bea
3(\alpha_q^2 + 1 -q^2)\left(\frac{2}{3}\alpha_q^2 + 1 - q^2\right)
\left(q^2 + \frac{\alpha_q^2}{2}\right) &=&
2\gamma_n q^2\alpha_q^2(2\gamma+\gamma_n q^2) \nonumber \\
\label{eq:eckhaus}
&-& \gamma_n q^2\left(2\gamma+\frac{3}{2}\gamma_n q^2\right)
\left(\frac{2}{3}\alpha_q^2 + 1 - q^2\right),
\eea
and the zig-zag stable region is given by $\alpha_q^2 > \frac{3}{2}(q^2-1)$.
Fig.~1 shows the Eckhaus and zig-zag stability boundaries for $\gamma=0.1$
(this value of $\gamma$ is motivated by recent experiments
\cite{re:tu89,re:ez86,re:ch92}) and $\gamma_n=0.05$. The reentrant shape
of the Eckhaus boundary is a direct consequence of the existence of a
small nonlinear damping coefficient $\gamma_{n}$. In the small
$\varepsilon$ limit ($\varepsilon \ll \gamma_{n}^{2}$), the region of
stability is symmetric around $q=1$ and is given by,
$\varepsilon > \frac{27}{32\gamma^2}(1-q^2)^2$, whereas for $\gamma_{n}
= 0$, the stable region is $\varepsilon <
\sqrt{\left[1+3(1-q^2)/(8\gamma)\right]^2} - 1$, which lies entirely in the
region $q > 1$. The parabolic stability boundary for $\varepsilon \ll
\gamma_{n}^{2}$ ($\gamma_{n} \neq 0$) can also be obtained from a
standing wave approximation to Eq. (\ref{eq:model_eq}), in analogy to
Milner's calculation \cite{re:mi91}. This approximation, however, fails
to reproduce Eq. (\ref{eq:eckhaus}) at larger values of $\varepsilon$.

Stability against transverse amplitude modulations can
be studied by assuming that $\psi(x,y,t) = [1+a(y,t)] \psi_0(x,q)$, with
$a(y,t) = a_0(t)\cos(Qy)$ ($Q$ small) and linearizing the resulting
equation for $a(y,t)$. The TAM
unstable region is given by,
\be
\label{eq:tam}
\frac{3}{16}(q^2-1)^2 > \frac{3}{4}\gamma_n^2\alpha_q^4
                        + \gamma\gamma_n\alpha_q^2,
\ee
and is shown in Fig.~1 as the shaded region. Instability occurs for
finite $Q$. If $\gamma_{n} = 0$, all roll solutions are unstable to TAM.
If, on the other hand, $\gamma_{n} \neq 0$, the region of wavenumbers
around $q=1$ is stable. At fixed $\gamma$, increasing $\gamma_{n}$
increases the width of this stable region.

We next turn to the results of our numerical calculation. We use a
pseudospectral method to solve Eq. (\ref{eq:model_eq}) on a square grid
of size $64 \pi ~ \times ~ 64 \pi$, with periodic boundary conditions.
The number of Fourier modes used for each axis is 256. Time stepping is
performed by a Crank-Nicholson scheme for the linear terms (including
$\psi^{*}$), and a second order Adams-Bashforth scheme for the nonlinear
terms. The time step used is $\Delta t =0.1$. The initial condition,
$\psi(x,y,t=0)$ is a set of Gaussianly distributed random numbers, of
zero mean and variance $10^{-4}$.
A typical run length is $10^{5} - 10^{6}$ time steps.

Fourier modes with wavenumber close to $q=1$ dominate in the early
linear regime. As the system enters the nonlinear regime, the circularly
averaged structure factor exhibits a dominant peak at $q = q_{max}$
which is seen to continuously shift away from 1 (detuning) and
towards the Eckhaus stable region for all values of $\varepsilon$,
i.e., the dominant
wavenumber of the roll pattern shifts to $q>1$. The value of $q_{max}$
at long times is shown by the circles in Fig. (1). The error bars
indicate the half-width at half-maximum of the peak. At $\varepsilon =
\varepsilon_{c}(\gamma_{n})$ the Eckhaus and TAM boundaries cross. For
$\varepsilon < \varepsilon_{c}$, there is a region of stability against
both Eckhaus and TAM perturbations. At sufficiently small $\varepsilon$,
an asymptotic stationary roll state is reached with $q_{max}$ inside
this region. For sufficiently large values of $\varepsilon$, $q_{max}$
enters the TAM unstable region first. Therefore the reentrant shape of
the Eckhaus boundary and the fact that it crosses the TAM boundary at
finite values
of $\varepsilon$ provides a mechanism for a finite threshold for the TAM
instability. The shift in $q_{max}$ and the position of the Eckhaus
and TAM boundaries depend strongly on the value of $\gamma_{n}$.

We finally describe the asymptotic temporal dependence of the
configurations as a function of $\varepsilon$. The results reported
correspond to $\gamma = 0.1$ and $\gamma_{n} = 0.05$, but they are
qualitatively similar in other cases provided that $\gamma_{n} \neq 0 $
(for larger $\gamma_{n}$, both the appearance of the TAM and the transition to
chaotic states are seen at larger $\varepsilon$).
An almost perfect and stationary roll pattern is found for $\varepsilon=0.02$
(Fig.~2(a)). As $\varepsilon$ is increased to $\varepsilon = 0.05$ a very
slowly varying transverse modulation of the roll pattern is observed
(Fig.~2(b)).
The wavelength of the modulation is about 3.2$\lambda_0$, with $\lambda_0=2\pi$
the critical wavelength at onset. The periodic and almost stationary
modulation suggests that the finite amplitude
modulated pattern at $\varepsilon=0.05$ can be described by $\psi(x,y)
= A_0(1+b_0\cos(Qy))\cos qx + \cdots$.
Fig.~3(b) shows the zeros of $\| \psi \|$ ($ \| \psi \| < \| \psi
\|_{max} /8 $ is considered a zero). They all originate from the $\cos qx$
factor. Hence the TAM is weak enough that $\|1+b_0\cos(Qy)\| > 0$.
At $\varepsilon=0.1$ (Fig.~2(c)), the modulation becomes stronger and
additional zeros of $\| \psi \|$ appear (Fig.~3(c)). We call the regions
in which the slowly varying component of $\| \psi \|$ is zero a
TAM defect. At this value of $\varepsilon$ the wave pattern becomes time
dependent, with the temporal variation of the wave pattern coming mostly
from the motion of TAM defects. At $\varepsilon=0.3$ (Fig.~2(d)), the length
scale of the modulation becomes smaller, and the density of
TAM defects increases (Fig.~3(d)). In order to characterize the temporal
fluctuations of the wave patterns in this state, we have calculated the power
spectrum of a time series of both
$\psi$ at a fixed point in space and the amplitude of a fixed Fourier mode.
Figure 4 shows the time series of $Re[\psi]$ at a fixed point (Fig.~4(a)), and
of the Fourier mode $q$=0.75 (Fig.~4(b)), and their corresponding power spectra
(Fig.~4(c) and (d)) for $\varepsilon=0.50$. The power spectra are broadband and
decay as a power law of frequency $\omega^{-z}$ with $z \approx 4.0$ for
frequencies in the range of $2 \times 10^{-3} < 2\omega/\Omega < 2 \times
10^{-2}$.  Similar power law decay of the power spectra is also observed
for $\varepsilon=0.3$. The power law decay and the value of the
exponent $z$ are in good agreement with experimental results \cite{re:tu89}.

We wish to thank Maxi San Miguel for many stimulating discussions, and for his
contributions to the understanding of the stability diagram. This work is
supported by the Microgravity Science and Applications
Division of the NASA under contract No. NAG3-1284.
This work is also supported in part by the Supercomputer
Computations Research Institute, which is partially funded by the U.S.
Department of Energy, contract No. DE-FC05-85ER25000.

\newpage

\begin{figure}
\label{fig:1}
\caption{Stability diagram for $\gamma=0.1$ and $\gamma_n=0.05$ in the
($q, \varepsilon$) plane: boundary of existence of stationary
roll solutions (dashed line, S); $q_{max}$, the peak of the circularly averaged
structure factor calculated numerically (circles with error bars indicating
the half-width at half-maximum of the peak); neutral stability curve (thin
solid line, N); Eckhaus stability boundary (thick solid line, E);
zig-zag stability boundary (long-dashed line, Z).
The TAM unstable region is the shaded area.}
\end{figure}

\begin{figure}
\label{fig:2}
\caption{Typical configurations of $Re[\psi]$ at long times
(shown in gray scale) for four values of $\varepsilon$. The initial
configuration was random.}
\end{figure}

\begin{figure}
\label{fi:3}
\caption{Zeros of $\| \psi \|$ ($ \| \psi \| < \| \psi \|_{max} /8 $ is
considered a zero) for four values of $\varepsilon$ at the same times as in
Fig. 2.}
\end{figure}

\begin{figure}
\label{fig:4}
\caption{(a) Time series of $Re[\psi]$ at a fixed point. (b) Time series of a
Fourier mode $Re \left[ \hat{\psi}(q_x = 0.75, q_y = 0) \right]$.
(c) Power spectrum of
the time series shown in (a). (d) Power spectrum of the time series shown
in (b). The dashed line represents a power law with an exponent of -4.}
\end{figure}

\end{document}